\documentstyle[preprint,aps]{revtex}

\date{\today}
\title{Manifestation of the Hofstadter butterfly in far-infrared absorption}
\author{Vidar Gudmundsson}
\address{Science Institute, University of Iceland, Dunhaga 3,
         IS-107 Reykjavik, Iceland.}
\author{Rolf R. Gerhardts}
\address{Max-Planck-Institut f\"ur Festk\"orperforschung, 
         Heisenbergstra{\ss}e 1,
         D-70569 Stuttgart,\\  Federal Republic of Germany}
\tighten

\begin{document}
\draft
\maketitle
\begin{abstract}
       The far-infrared absorption of a 
       two-dimensional electron gas with a square-lattice modulation 
       in a perpendicular constant
       magnetic field is calculated self-consistently within the 
       Hartree approximation. 
       For strong modulation and short period
       we obtain intra- and intersubband magnetoplasmon modes
       reflecting the subbands of the Hofstadter butterfly
       in two or more Landau bands. The character of the 
       absorption and the correlation of the peaks to
       the number of flux quanta through each
       unit cell of the periodic potential depends strongly 
       on the location of the chemical potential with respect to
       the subbands, or what is the same, on the density of 
       electrons in the system. 
\end{abstract}
\pacs{71.20.-b 73.20.Dx 73.20.Mf 78.66.Fd}
%

%
Indications of the Hofstadter subband 
structure\cite{Hofstadter76:2239,Pfannkuche92:12606} of the Landau bands
of a two-dimensional electron gas (2DEG) in square
modulated lateral superlattices have been believed to be found 
directly\cite{Schloesser96:683} and indirectly\cite{Gerhardts91:5192}
in transport measurements. This opens the important question of whether the
subband structure can directly be detected in far-infrared (FIR) absorption
measurements. Calculations of the ground state properties of the 
laterally modulated 2DEG indicate that the strong screening effects
would hamper an observation of even the coarse structure of the
Hofstadter butterfly in, e.g.\ magnetocapacitance or magnetoresistance
measurements, unless very short periods and strong modulation
would be used\cite{Gudmundsson95:16744}. But what about FIR measurements? 

The FIR-absorption is determined by the self-consistent field which
describes the dynamical screening of the incident external field.
Since, as calculations \cite{Gudmundsson95:16744} show, screening 
is also very important in the ground state, screening effects should be  
treated on equal footing in both the ground state and in the FIR response.
That an inconsistent treatment of screening effects may lead to
qualitatively wrong results in a 2DEG with strong lateral modulation
is known from the case of quantum dots. There, according to the 
generalized Kohn's theorem\cite{Maksym90:108,Kohn61:1242}, 
Coulomb interaction effects on the ground
state and dynamical response of a parabolically confined electron system
cancel, so that the dot responds at the bare single-electron frequencies. 
This experimentally confirmed fact has been reproduced only by those 
many-body calculations which have treated the interaction effects in 
the ground state and in the FIR response on the same footing. Results of such
calculations have been published for isolated quantum 
dots\cite{Gudmundsson91:12098,Pfannkuche91:13132} 
and for unidirectionally modulated 2DEG\cite{Wulf90:3113,Wulf90:7637},
but to our knowledge no corresponding results are available for the 
square modulation.\footnote{The collective excitations of a 2DEG with
a two-dimensional magnetic-field modulation have been studied by
X.\ Wu and S.\ E.\ Ulloa in Phys.\ Rev.\ B {\bf 47}, 10028 (1993), but 
without considering interaction effects in the ground state.} 

%
%
%

The square lateral superlattice with period $L$ is spanned 
by the lattice vectors
${\bf R}=m{\bf l}_1+n{\bf l}_2$, where ${\bf l}_1=L\hat{\bf x}$, and
${\bf l}_2=L\hat{\bf y}$ are the primitive translations of the Bravais
lattice ${\cal B}$; $n,m\in Z$. The reciprocal lattice ${\cal R}$ is spanned by
${\bf G}=G_1{\bf g}_1+G_2{\bf g}_2$, with ${\bf g}_1=2\pi \hat{\bf y}/l_1$,
${\bf g}_2=2\pi \hat{\bf x}/l_2$, and $G_1,G_2\in Z$. 
The ground-state properties 
of the interacting 2DEG in a perpendicular homogeneous magnetic field
${\bf B}=B\hat{\bf z}$ and the periodic potential
$V({\bf r})=V_0\{\cos (g_1x)+\cos (g_2y)\}$ are calculated in the
Hartree approximation. The Hartree single electron states $|\alpha )$
and their energies $\varepsilon_{\alpha}$ are labelled by the quantum
numbers $\{n_l,\mu ,\nu\}=\alpha$, where $n_l\geq 0$ is a Landau band index,
$\mu =(\theta_1+2\pi n_1)/p$, $\nu =(\theta_2+2\pi n_2)/q$, with
$n_1\in I_1=\{0,\cdots p-1\}$, $n_2\in I_2=\{0,\cdots q-1\}$,
$\theta_i\in [-\pi ,\pi ]$, and $pq\in N$  
is the number of magnetic flux quanta through the lattice unit cell. 
The Hartree states are expanded in terms of the symmetric basis functions
constructed by Ferrari\cite{Ferrari90:4598} and used by 
Silberbauer\cite{Silberbauer92:7355} and the present 
authors\cite{Gudmundsson95:16744}. 
The Ferrari basis is complete provided that $(\mu ,\nu )\neq (\pi ,\pi)$ 
for all $(n_1,n_2)\in I_1\times I_2$. 
The Hartree potential ``felt'' by each
electron and caused by the total electronic charge density of 
the 2DEG, $-en_s({\bf r})$,
and the positive neutralizing background charge, $+en_b$, together with
the external periodic potential $V({\bf r})$ do not couple states at 
different points in the quasi-Brillouin zone 
${\bf\theta}=(\theta_1 ,\theta_2 )\in\{ [-\pi ,\pi]\times [-\pi ,\pi]\}$.
However, states at a given point depend,
in a self-consistent way, on states in the whole zone
through $n_s({\bf r})$\cite{Silberbauer92:7355,Gudmundsson95:16744}.
The periodic potential broadens the Landau levels into Landau bands that due
to the comensurability conditions between $L$ and the magnetic length
$l=(c\hbar /eB)^{1/2}$, are split into $pq$ 
subbands. The resulting energy spectrum, the Hofstadter 
butterfly\cite{Hofstadter76:2239,Pfannkuche92:12606}, retains essentially it's
complicated gap structure but is strongly reduced in symmetry by the
electron-electron interaction\cite{Gudmundsson95:16744} 
and coupling to higher Landau bands
caused by the periodic potential\cite{Petschel93:239}. The symmetry
depends on the mean electron density, or, equivalently, the chemical potential.

In order to calculate the FIR absorption of the 2DEG we perturb this by
a monochromatic external electric field:
\begin{equation}
      {\bf E}_{ext}({\bf r},t)=-i{\cal E}_0\frac{{\bf k}+{\bf G}}
      {|{\bf k}+{\bf G}|}\exp{\left\{ i({\bf k}+{\bf G})\cdot{\bf r}
      -i\omega t\right\} } .
\label{Eext} 
\end{equation}
Here we do not restrict the
dispersion relation for the external field, $\omega({\bf k}+{\bf G})$,
to that of a free propagating electromagnetic
wave, instead we allow for the more general situation in which the external
field is produced as in a near-field spectroscopy or in a 
Raman scattering set-up. 
The power absorption is found from the Joule heating of the self-consistent
electric field\cite{Dahl90:5763,Gudmundsson91:12098}, 
$-{\bf\nabla}\phi_{sc}$, with $\phi_{sc}=\phi_{ext}+\phi_{ind}$,
\begin{equation}
      P({\bf k}+{\bf G},\omega )=-\frac{\omega}{4\pi}\left[ 
      |{\bf k}+{\bf G}|\phi_{sc}({\bf k}+{\bf G},\omega )
      \phi_{ext}^*({\bf k}+{\bf G},\omega )\right] .
\label{Pom} 
\end{equation}
The induced potential $\phi_{ind}$ is caused by the density 
variation $\delta n_s({\bf r})$ due to $\phi_{sc}$, 
which can then in turn be related to the
external field by the dielectric matrix
\begin{equation}
      \sum_{{\bf G}'}\epsilon_{{\bf G},{\bf G}'}({\bf k},\omega )
      \phi_{sc}({\bf k}+{\bf G}',\omega )=\phi_{ext}({\bf k}+{\bf G},\omega ) .
\label{epsilon} 
\end{equation}
The dielectric tensor, $\epsilon_{{\bf G},{\bf G}'}({\bf k},\omega )
=\delta_{{\bf G},{\bf G}'}-\frac{2\pi e^2}
{\kappa |{\bf k}+{\bf G}|}\chi_{{\bf G},{\bf G}'}({\bf k},\omega )$,
is determined by the susceptibility of the 2DEG 
\begin{equation}
\begin{array}{rl}
      \chi_{{\bf G},{\bf G}'}({\bf k},\omega )=\frac{1}
      {(2\pi L)^2}\int d{\bf \theta}
      \sum_{\alpha ,\alpha '}f_{\alpha ,\alpha '}
      ^{{\bf\theta},{\bf\theta}-{\bf k}L}(\hbar\omega )
      & J_{\alpha ,\alpha '}^{{\bf\theta},{\bf\theta}-{\bf k}L-{\bf G}L}
      ({\bf k}+{\bf G})\\     { }   &
      \left(J_{\alpha ,\alpha '}^{{\bf\theta},{\bf\theta}-{\bf k}L-{\bf G}L}
      ({\bf k}+{\bf G}')\right) ^* , 
\end{array}
\label{chi}
\end{equation}
where ${\bf k}$ is in the first Brillouin zone, $\kappa$ is the dielectric
constant of the surrounding medium, 
\begin{equation}
      f_{\alpha ,\alpha '}
      ^{{\bf\theta},{\bf\theta}'}(\hbar\omega )=
      \left\{ \frac{f^0(\varepsilon_{\alpha ,{\bf\theta}})-
      f^0(\varepsilon_{\alpha ' ,{\bf\theta}'})}
      {\hbar\omega+(\varepsilon_{\alpha ,  {\bf\theta}}-
                    \varepsilon_{\alpha ' ,{\bf\theta}'})+i\hbar\eta } 
                                                                \right\} 
\label{faa}
\end{equation}
in which $f^0$ is the equilibrium Fermi distribution, $\eta\rightarrow 0^+$, 
and 
\begin{equation}
           J_{\alpha ,\alpha '}^{{\bf\theta},{\bf\theta}'}({\bf k})=
           (\alpha '({\bf\theta}')|e^{-i{\bf k}\cdot{\bf r}}|\alpha (\theta)) .
\end{equation}
Special care must be taken with respect to the symmetry 
of the the wave functions corresponding to the Hartree states 
$|\alpha{({\bf\theta}}))$ when translating them across the boundaries 
of the quasi-Brillouin zones. The peaks in the absorption spectra represent 
absorption of collective modes of the 2DEG. Single-electron transitions
and collective oscillations can be identified as zeros of the 
real part of the determinant of the dielectric matrix. For the
collective oscillations the imaginary part simultaneously vanishes but
becomes very large for the single electron transitions indicating the
strong damping of such processes. Here we prefer to calculate 
$P({\bf k}+{\bf G},\omega )$ for a monochromatic external field
instead of searching for zeros of  
$\det \{\epsilon_{{\bf G},{\bf G}'}\}$ in order to avoid 
the complication of multiple branches of the magnetoplasmon 
dispersion for different reciprocal lattice vectors ${\bf G}$.
The response of the system can also be analyzed in more detail
by imposing a certain symmetry or polarization on the external field. 
Although we study collective oscillations here, it is still 
possible to identify in many cases the most important 
single-electron transitions contributing to the collective mode, either 
from the dispersion or by restricting the type of single electron
transitions in the calculation. 

%
%
In the numerical calulations we use GaAs parameters, $m^*=0.067m_0$, and
$\kappa =12.4$. The power dissipation of the 2DEG is made possible by 
retaining a small but finite imaginary part for the frequency in the
susceptibility, Eqs.\ (\ref{chi}) and (\ref{faa}).
We choose $\eta =\omega_c/50$. 
In order to examine the fine structure of the magnetoplasmon
dispersion we have chosen a monochromatic external perturbing 
field (\ref{Eext}) with ${\bf G}=0$ and ${\bf k}$ in the first Brillouin
zone. 

The absorption of a homogeneous 2DEG ($V_0=0$) is compared in 
Fig.\ \ref{Bernstein} with the first-order dispersion in ${\bf k}$
of a magnetoplasmon,
$\omega^2=\omega_c^2+2\pi e^2n_sk/(\kappa m^*)$. 
Kohn's theorem manifests itself by the fact that
only one peak is visible for ${\bf k}\rightarrow 0$\cite{Kohn61:1242}, 
and for finite wave vectors ${\bf k}$ the magnetoplasmon 
interacts with harmonics of the
cylcotron resonance, resulting in so the called Bernstein 
modes\cite{Bernstein58:10,Horing76:216,Batke86:6951}.

To make the Hofstadter subband structure of the Landau bands discernible,
we have to resort to short periods $L=50\:$nm and strong modulation 
$V_0=4\:$meV, just as we had to in order to observe the subbands in the
thermodynamic density of states\cite{Gudmundsson95:16744}. In addition,
the structures are very sensitive to the location of the chemical
potential with respect to the subbands, i.e.\ to the filling factor
$\nu$ of the Landau bands. Figure\ \ref{Intra}a shows the absorption for
$\nu =1/2$, and the low frequency part is reproduced in Fig.\ \ref{Intra}b.
In Fig.\ \ref{Intra}c the subband dispersion in one 
quasi-Brillouin zone along several values of $\theta_2$ 
is projected onto the $\theta_1$-axis, 
thus indicating the bandwidths and the position
of the chemical potential. Since one and a half subband in the lowest
Landau band are filled, and it is not inhibited by a selection rule,
we can see {\sl three intraband} magnetoplasmons. 
The $k$-dispersion allows the following identification in Fig.\ \ref{Intra}b:
The lowest frequency peak is the plasmon with contributions from
transitions between the upper two subbands, the center peak gets
contributions from the two lowest subbands and the high frequency
peak has contributions connecting the lowest and the top subband.
Since only two subbands of the lowest Landau band are occupied
(of which one is partially filled), only
two peaks are found for the {\sl interband} magnetoplasmon with
frequency just above $\omega_c$. 
The absorption for the same situation
but for a different filling factor, i.e.\ $\nu =5/6$, 
is seen in Fig.\ \ref{Inter}.
Now all the three subbands of the lowest Landau band are at least
partially occupied and {\sl three interband} magnetoplasmon absorption
peaks are clearly discernible. For $\nu =1$ the low frequency intraband
peaks vanish and the interband peak with the lowest frequency accounts for
almost all of the oscillator strength. 
The structure of the Hofstadter spectrum can be seen for 
other values of $pq$ in the calculations. In Fig.\ \ref{pq24}
the absorption for the cases $pq=2$ and $pq=4$ is seen for quite
different parameters. The double structure expected for $pq=2$ is visible
for relatively large period, $L=100\:$nm, and weak modulation
$V_0=0.4\:$meV, while two of the four peaks for the case of $pq=4$ are just
appearing for $V_0=3\:$meV, and $L=50\:$nm. This reflects the 
stronger screening of the 2DEG observed in the ground state
calculations\cite{Gudmundsson95:16744} for higher values of
$pq$, where a strong modulation and short period were needed to see
structure in the thermodynamic density of states caused by the
Hofstadter subbands. 
In addition, here the filling factor $\nu$ not only determines the 
strength of the screening and thus the relevance of selection rules, 
but it is also essential to control in order to occupy subbands such
that single electron transitions can take place contributing to
the collective oscillations.

It is interesting to note that
in the case of a nonvanishing potential modulation, $V_0\neq 0$, 
the absorption $P(\omega )$ has a nontrivial structure, even for the
case ${\bf k}\rightarrow 0$. 
The condition for the validity of the  
generlized Kohn's theorem can thus be violated by
either spatially modulating the time-dependent external electric 
field (\ref{Eext}) or by the static potential $V({\bf r})$ the 2DEG
resides in.

In summary, the FIR-absorption of a 2DEG in a short-period strongly modulated 
square lateral superlattice shows the underlying Hofstadter 
subband structure. The structure can be seen in the absorption
due to either intra- or interband magnetoplasmons by tuning the density or the
filling factor within the lowest Landau band. 
The intraband magnetoplasmon peaks in the absorption have weak oscillator 
strengths compared to the interband absorption and are situated in a region
on the energy scale that is difficult to detect with present day 
FIR technology. 

\acknowledgements{We thank Dr.\ B.\ Farid for an interesting discussion.  
This research was supported in part by the Icelandic
Natural Science Foundation, the University of Iceland Research Fund,
and a and a NATO collaborative research Grant No. CRG 921204.}
\bibliographystyle{prsty}

%
%
%
\begin{figure}
\caption{The power absorption $P(k_1,\omega )$ for a homogeneous
         2DEG ($V_0=0$). The dashed straight lines represent the frequencies
         of the cyclotron resonances $\omega =n\omega_c$ and the 
         dashed curve is the dispersion of the magnetoplasmon in
         first order with respect to $k_1$. $L=200\:$ nm, $pq=1$, $\nu =1$, 
         $m^*=0.067m_0$, $T=1\:$K, $\kappa =12.4$, and 
         $\hbar\omega_c=0.178\:$meV.} 
\label{Bernstein}
\end{figure}
\begin{figure}
\caption{(a) The power absorption $P(k_1,\omega )$ (in arbitrary units) 
         for a modulated 
         2DEG ($V_0=4\:$meV). The solid curve in the ($k_1,\omega$)-plane
         represents the dispersion of the magnetoplasmon 
         of a homogeneous 2DEG in first order with respect to $k_1$. 
         (b) The low frequency region repeated
         from (a). The left peak decreases with
         increasing wave vector, $k_1$. 
         (c) The dispersion of the two 
         lowest Landau bands in the quasi-Brillouin zone 
         along several values of $\theta_2$ projected on the
         $\theta_1$ axis (each band
         is split into three subbands). The chemical potential is
         indicated by a solid horizontal line.
         $pq=3$, $L=50\:$nm, $\nu =1/2$, and $\hbar\omega_c=8.57\:$meV.}
\label{Intra}
\end{figure}
\begin{figure}
\caption{(a) The power absoprtion $P(k_1,\omega )$ for a modulated 
         2DEG ($V_0=4\:$meV). The solid curve in the ($k_1,\omega$)-plane
         represents the dispersion of the magnetoplasmon in
         a homogeneous 2DEG in the first order with respect to $k_1$.
         (b) The dispersion of the two
         lowest Landau bands in the quasi-Brillouin zone 
         along several values of $\theta_2$ projected on the
         $\theta_1$ axis (each band
         is split into three subbands). The chemical potential is
         indicated by a solid horizontal line.
         $pq=3$, $L=50\:$nm, $\nu =5/6$, and $\hbar\omega_c=8.57\:$meV.}
\label{Inter}
\end{figure}
\begin{figure}
\caption{The power absorption $P(k_1,\omega )$ for a modulated 
         2DEG in two cases; (solid) $pq=2$, $L=100\:$nm, $\nu =1/2$,  
         $\hbar\omega_c=1.43\:$meV, and $V_0=0.4\:$meV, and
         (dashed) $pq=4$, $L=50\:$nm, $\nu =7/8$,  
         $\hbar\omega_c=11.4\:$meV, and $V_0=3\:$meV. 
         $k_1L=0.6$, and $T=1\:$K in both cases. }
\label{pq24}
\end{figure}
%
\end{document}